\documentstyle[prl,preprint,multicol,aps]{revtex} \input epsf

\begin{document}

\draft 

\title{Inherent Structure Approach to the Study of Glass Forming Liquids} 

\author{Srikanth Sastry$^{*}$}
\address{Jawaharlal Nehru Centre for Advanced Scientific Research, 
Jakkur Campus, Bangalore 560064, INDIA}                            

\maketitle

\begin{abstract}
The inherent structure approach, wherein thermodynamic and structural
changes in glass forming liquids are analyzed in terms of local
potential energy minima that the liquid samples, has recently been
applied extensively to the study of thermodynamic aspects of glass
forming liquids. The evalation of the configurational entropy, which
arises from the multiplicity of local energy minima, plays a central
role in such analyses. Results are presented here concerning the
calculation of configurational entropy based on computer simulations
of a model liquid, and the application of the inherent structure
formalism to the study of the glass transition locus, and the fragility 
of glass forming liquids. 
\end{abstract} 

%\pacs{PACS numbers:64.70.Pf, 64.60.My, 64.90.+b,61.20.Lc,63.50.+x}

%\narrowtext 

%\begin{multicols}{2}

\section{Introduction} 

Most liquids, when cooled to low temperatures in such a fashion that
the equilibrium crystal phase does not form, undergo a lower
temperature transformation to a solid phase which is {\it amorphous},
{i.e.}, lacking the long range periodic structure that characterizes
crystalline solids. The amorphous phase is a {\it glass}, and the
transformation process is referred to as the glass transition. In the
process of cooling towards the glassy state, liquids display a rapid
increase in viscosity, and display other unusual dynamical features
such as stretched exponential relaxation and heterogeneous
dynamics\cite{science,debene,ediger,angellrev,trieste}. The glass
transition observed in experiments (the ``laboratory glass
transition'') is understood to be not a transition in the
thermodynamic sense but a falling out of equilibrium of the
liquid. Whether a thermodynamic transition underlies the laboratory
transition, and how an explanation of such a thermodynamic transition
is linked (or not) to a detailed understanding of the unusual dynamics
displayed by liquids at low temperatures, is a topic is considerable
current research\cite{trieste}. A particular approach that has been
studied, based on the analysis of local minima of the liquid's
potential energy (termed ``inherent structures'')\cite{inh} is
described here.  Section \ref{secinh} outlines the approach based the
study of inherent structures. Section \ref{secsim} describes the
details of computer simulations of a model liquid employed in
implementing the analysis of some properties of glass forming liquids, 
along with the results. Section \ref{secconc} contains a summary of 
the results and conclusions. 

\section{The Inherent Structure Approach}\label{secinh}

The disordered structure of a liquid has the implication that the
energies of interaction between particles will generally be very
complicated, and that the part of configuration space explored by the
material in the liquid state is characterized by the presence of many
local minima of the potential energy. Such is the case also, {\it
e. g.}, in a crystalline solid if one allows for the presence of
defects. The use of the phrase ``energy landscape'' to describe the
complicated interactions in a glass forming liquid (and other
disordered systems) therefore contains in addition the expectation
that the complicated potential energy topogrphy plays an essential
role in determining the properties of the system. If such is the case,
it is desirable to attempt a description of glass forming liquids in
terms of quantities that define the nature of the potential energy
landscape. In the inherent structure approach, one considers the
decomposition of the $3 N$ dimensional (for an atomic liquid)
configuration space of the liquid into basins of individual local
potential energy minima. A basin of a given minimum is defined as the
set of points in the configuration space (or configurations) which map
to that minimum under a local energy minimization. The canonical
partition function of the liquid can then be expressed as a sum over
inherent structure basins, the summand being partial partition
functions defined for individual basins. In the following, the equations are
written for a two-component atomic liquid, since the model liquid
described in the next section is such a liquid. In turn, the sum over
basins is written in terms of (a) a distribution of minima in energy, 
and (b) the free energies of basins, as follows:  
\begin{eqnarray}\label{eq:pfn}
Q(N,\rho,T) = \Lambda^{-3N} {1\over N_A! N_B!} \int d{\bf r}^N exp\left[-\beta \Phi\right]\\
	    = \sum_\alpha exp\left[-\beta \Phi_\alpha\right] \Lambda^{-3N} \int_{V_\alpha} d{\bf r}^N exp\left(-\beta(\Phi-\Phi_\alpha)\right)\nonumber \\
	    = \int d\Phi ~ \Omega(\Phi) ~exp\left[-\beta(\Phi + F_{vib}(\Phi,T))\right]\nonumber \\
	    = \int d\Phi ~exp\left[-\beta (\Phi + F_{vib}(\Phi,T) - T {\cal S}_c(\Phi))\right]\nonumber
\end{eqnarray}
where $\Phi$ is the total potential energy of the system, $\alpha$
indexes individual inherent structures, $\Phi_\alpha$ is the potential
energy at the minimum, $\Omega(\Phi)$ is the number density of
inherent structures with energy $\Phi$, and the configurational
entropy density ${\cal S}_c \equiv k_B \ln \Omega$. The {\it basin free
energy} $F_{vib}(\Phi_\alpha,T)$ is obtained by a restricted partition
function sum over a given inherent structure basin, $V_\alpha$.
$\Lambda$ is the de Broglie wavelength, $N_A$ and $N_B$ are the number
of $A$ and $B$ type atoms in the two component liquid, $T$ is the
temperature, and $\rho$ the density of the liquid. In the following, the 
dependence on $\rho$ is not explicitly stated always since the interest
is in $T$ dependent behavior at constant density. The configurational
entropy of the liquid arises from the multiplicity of local potential
energy minima sampled by the liquid at temperature $T$, and is related
to the configurational entropy density above by
\begin{equation}\label{eq:scdos}
S_c(T) = \int d\Phi~ {\cal S}_c(\Phi) P(\Phi,T),\\
\end{equation}
where 
\begin{eqnarray}\label{eq:pdef}
P(\Phi,T) = \Omega(\Phi) ~exp\left[-\beta (\Phi + F_{vib}(\Phi,T))\right]/Q(N,\rho,T),\\
	  = exp\left[-\beta (\Phi + F_{vib}(\Phi,T) - T {\cal S}_c(\Phi))\right]/Q(N,\rho,T),
\nonumber
\end{eqnarray} 
is the probability density that inherent structures
of energy $\Phi$ are sampled at temperature $T$. In the above
expression for the partition function, an assumption has been made
that the basin free energy does not differ for different basins of the
same inherent structure energy. Without reference to the distribution of 
minima, the configurational entropy can be defined as the difference of 
the total entropy of the liquid and the vibrational entropy of typical 
minima sampled at a given temperature: 
\begin{equation} \label{eq:scdef}
S_c(\rho,T) = S_{total}(\rho,T) - S_{vib}(\rho,T).
\end{equation}

	The ``entropy theory'' of Gibbs, Di Marzio and
Adam\cite{dimarz,adam-gibbs} define the ideal glass transition,
underlying the laboratory transition, as an ``entropy vanishing''
transition where the configurational entropy vanishes (the
configurational entropy is not, however, defined in precisely the same
way in \cite{dimarz,adam-gibbs}). A similar picture also emerges from
the study of mean field spin glass models and calculations motivated
by them\cite{kirk,parmez,silvio}.  Whether such a transition exists
for real materials is still a matter of
debate\cite{gujrati,FHS88}. The calculations below produce such an
entropy vanishing transition but it must be kept in mind that they
result from extrapolations which may not be valid.

	Further, Adam and Gibbs\cite{adam-gibbs} theory relates the
configurational entropy to relaxation times in the liquid:
\begin{equation}  \label{eq:ag}
\tau = \tau_0 \exp\left[{A\over T S_c}\right], 
\end{equation}
where $A$ is a material specific constant. The validity of this
relation has been verified by numerous experimental studies (which
typically use the {\it excess} entropy of the liquid over the crystal
in place of $S_c$) and computer simulation
studies\cite{speedyag,scala,sastryprl,sastrynat2} (where
configurational entropy is evaluated). Further, if $S_c$ has the form
$T S_c = K_{\small AG}(T/T_K - 1)$, the Adam-Gibbs relation results in
the Vogel-Fulcher-Tammann-Hesse (VFT) form, observed to describe the $T$
dependence of viscosity, as well as diffusivity and relaxation times,
in many glass formers. The VFT relation may be written as
\begin{equation} \label{eq:vft}
\tau = \tau_0 \exp\left[{1\over K_{\small VFT}(T/T_0 - 1)}\right], 
\end{equation}
where $T_0$ is the temperature of apparent divergence of
viscosity. $K_{\small VFT}$ is a material specific parameter
quantifying the {\it kinetic} fragility. Fragility is a measure of how
rapidly the viscosity, relaxation times, {\it etc} of a liquid changes
as the glassy state is approached\cite{fragility}. 

That the basin free energy $F_{vib}$ arises from ``vibrational''
motion within individual basins is emphasized by the suffix {\it
vib}. If this motion is sufficiently localized around the minima, a
suitable procedure would be to approximate the basins as harmonic
wells, and to evaluate the basin free energy within this
approximation. The validity of such a procedure has been tested
recently in various
studies\cite{fs,heuer,parmez,sastryprl,sastrynat2,sastrypcc}. It
is found that below the temperature where the liquid begins to exhibit
aspects of {\it slow dynamics}, (non-Arrhenius behaviour of relaxation
times, and stretched exponential
relaxation)\cite{sastrynature,fs,sastrypcc}, a harmonic approximation
of the basins is reasonable. Additional evidence is provided in this
paper concerning the validity of this approximation, as well as a
classical (as opposed to quantum mechanical) treatment of the basin
partition function. A classical calculation of the basin free energy 
yields
\begin{equation}\label{eq:fbasin}
F_{vib}  = k_B T \sum_{i = 1}^{3N} \ln {h \nu_i \over k_B T},
\end{equation}
or equivalently, the basin entropy,  
\begin{equation} \label{eq:sbdef}
{S_{vib} \over k_B} =  \sum_{i = 1}^{3N} 1 - \log({h \nu_i \over  k_B T}),
\end{equation}
where $\nu_i$ are the vibrational frequencies of the given basin, and
$h$ is Plank's constant. From the form of $S_{vib}$ it is apparent
that the entropy difference between two basins arises solely due to
the difference in their frequencies. Thus, such entropy differences
remain finite as $T \rightarrow 0$ which is unphysical as the basin
entropy of each basin and therefore their difference must go to zero
for $T = 0$. The extent to which this artifact affects the
applicability of classical calculations employed in many recent
studies depends on the deviation of the quantum mechanical result at
the glass transition temperature (since below the glass transition,
the system occupies a single basin). The quantum mechanical basin free energy
is given, in terms of the same vibrational frequencies $\nu_i$, as\cite{ashcroft}
\begin{equation}\label{eq:fbasinq}
F_{vib}  = k_B T \sum_i^{3N} \ln \left[\exp\left({h \nu_i \over 2 k_B T}\right) - \exp\left({-h \nu_i \over 2 k_B T}\right)\right],
\end{equation}
and the basin entropy as
\begin{equation} \label{eq:sbdefq}
{S_{vib} \over k_B} = \sum_{i = 1}^{3N} {h \nu_i \over k_B T} \left[{1 \over 2} + {1 \over \exp\left({h \nu_i \over k_B T}\right) - 1}\right] - \sum_{i = 1}^{3N} \ln \left[\exp\left({h \nu_i \over 2 k_B T}\right) - \exp\left({-h \nu_i \over 2 k_B T}\right)\right].
\end{equation}
A comparison will be made in the next section of these two
(Eq.(\ref{eq:sbdef}) and Eq.(\ref{eq:sbdefq})) expressions for
$S_{vib}$. Calculations based on Eq.(\ref{eq:sbdef}), where the
vibrational frequencies are obtained numerically for energy minima
generated in simulations, indicate\cite{sastrynat2} (see also
\cite{fsaging,water}) that the difference in $S_{vib}$, between basins
is roughly linear in the basin energy. Thus one can write
\begin{equation} 
\Delta S_{vib}(\Phi) \equiv S_{vib}(\Phi,T) - S_{vib}(\Phi_0,T) = \delta S~ (\Phi - \Phi_0),
\end{equation} 
and correspondingly,
\begin{equation} \label{eq:fvib}
F_{vib}(\Phi,T) = F_{vib}(\Phi_0,T) - T \delta S (\Phi - \Phi_0)
\end{equation}
where $\Phi_0$ is a reference basin energy. The latter expression follows 
since the internal energy $U_{vib} = 3 N k_B T$ for all basins. 

	In addition to the basin free energy, the partition function
in Eq.(\ref{eq:pfn}) requires knowledge of the configurational entropy
density ${\cal S}_c$. Various recent studies have explored methods for
estimating ${\cal S}_c$ from computer
simulations\cite{speedydist,fs,heuer,sastrytrieste,sastrynat2,crisanti}. It has
been observed that the distribution $\Omega(\Phi)$ is well described
by a Gaussian\cite{speedydist,heuertrieste,sastrynat2} (equivalently,
${\cal S}_c(\Phi)$ an inverted parabola). Although the
arguments\cite{speedydist,heuertrieste} may not apply to low energy
minima, a Gaussian form for $\Omega(\Phi)$ allows for a
straightforward evaluation of the partition function
Eq.(\ref{eq:pfn}), and whose validity has been tested in the range
of temperatures where simulations are performed\cite{sastrynat2,sastrypcc}. 
The configurational entropy density is written as 
\begin{equation}\label{eq:scdos2} 
{{\cal S}_c(\Phi) \over N k_B} = \alpha - {(\Phi - \Phi_o)^2\over \sigma^2}
\end{equation} 
where $\alpha$ is the height of the parabola and determines the total
number of configurational states, {\it i. e.} energy minima (the total
number is proportional to $\exp(\alpha N)$ ), $\Phi_0$ and $\sigma^2$
respectively define the mean and the variance of the distribution.
The parameters $\alpha$, $\Phi_0$ and $\sigma$ have been estimated
from simulation data\cite{sastrynat2}. With the above form for ${\cal
S}_c(\Phi)$ and Eq.(\ref{eq:fvib}) for the vibrational free energy,
the partition function can be evaluated, from which the following
temperature dependence of the configurational entropy, the ideal glass
transition temperature $T_K$ (defined by $S_c(T_K) = 0$ and the
inherent structure energies are obtained:
\begin{equation} \label{eq:phi}
<\Phi> (T) = \Phi_0^{eff} - {\sigma^2 \over 2 N k_B T},
\end{equation} 
where $\Phi_0^{eff} = \Phi_0 + {\sigma^2 \delta S \over 2 N k_B}$,  
\begin{equation} \label{eq:kagPEL}
T S_c(T) = K^{\tiny{PEL}}_{\small AG}(T)~~ (T/T_K -1); ~~~K^{\small{\small PEL}}_{\small AG}(T) = \left({\sigma \sqrt{\alpha} \over 2} + {\sigma^2 \delta S\over 4 N k_B}\right) \left( 1 + {T_K\over T}\right) - {\sigma^2 \delta S\over 2 N k_B},
\end{equation}
and
\begin{equation}
T_K = \sigma(2 N k_B\sqrt\alpha + \sigma \delta S)^{-1}.
\end{equation}
These equations constitute relations that express quantities relevant
to the thermodynamics of glass forming liquids, the configurational
entropy and the ideal glass transition temperature, in terms of
paramters that describe the ``energy landscape'' of the liquid, namely
the distribution of local energy minima, and the topography of
individual minima in the form of vibrational frequencies. In
particular, the expressions for $T S_c$ shows that the fragility of
the liquid can be expressed in terms of parameters that quantify the
``energy landscape'' of the liquid.

\section{Simulation Details and Results}\label{secsim}

Many recent computer simulation studies of dynamics and thermodynamics
aspects of glass forming liquids have employed a binary mixture of
atomic particles as a model
system\cite{kob,kobhet,sastrynature,fs,parmez,sastryprl}, as this
system has been parametrized to prevent crystallization.  Results
presented here are from molecular dynamics simulations of $204$ type
$A$ and $52$ type $B$ particles. The particles interact {\it via} the
Lennard-Jones (LJ) potential, with parameters
$\epsilon_{AB}/\epsilon_{AA} = 1.5$, $\epsilon_{BB}/\epsilon_{AA} =
0.5$, $\sigma_{AB}/\sigma_{AA} = 0.8$, and $\sigma_{BB}/\sigma_{AA} =
0.88$, and $m_B/m_A = 1$. The LJ potential is modified with a
quadratic cutoff and shifting at $r_c^{\alpha \beta} = 2.5
\sigma_{\alpha \beta}$\cite{stoddard}.  All quantities are reported in
reduced units, length in units of $\sigma_{AA}$, volume $V$ in units
of $\sigma_{AA}^{3}$ (density $\rho \equiv N/V$, where $N$ is the
number of particles, in units of $\sigma_{AA}^{-3} \equiv \rho_0$),
temperature in units of $\epsilon_{AA}/k_B$, energy in units of
$\epsilon_{AA}$ and time in units of $\tau_m \equiv (\sigma_{AA}^2
m/\epsilon_{AA})^{1/2}$, where $m = m_A = m_B$ is the mass of the
particles. Argon units are used for the $A$ type particles when it is
desirable to state values in SI units, {\it i. e.}  $\epsilon_{AA} = 119.8 K
\times k_B$, $\sigma_{AA} = 0.3405 ~nm$, $m_A = 6.6337 \times 10^{-26}
~kg$. Molecular dynamics simulations are performed over a wide
range of temperatures at each density, with run lengths ranging from 
$1.3 ns$ to $0.4 \mu s$. Local energy minimizations
are performed for $1000$ ($k_B T/\epsilon_{AA} < 1.$) or $100$ ($k_B
T/\epsilon_{AA} > 1.$) configurations to obtain typical local energy
minima or `inherent structures'\cite{inh} sampled by the liquid. The
Hessian (matrix of second derivatives of the potential energy)
evaluated at the minima are diagonalized to obtain the vibrational
frequencies. 

	As the binary LJ liquid has mostly been studied at low
temperatures and high densities, the location of the liquid-gas
critical point has not previously been estimated. A rough estimate is
obtained here by performing simulations in the appropriate range of
$\rho$ and $T$ of length $2.912 ns$. Isotherms close to the critical
point are shown in Fig. 1, which result in an estimate of the
liquid-gas critical point of $k_B T_c/\epsilon_{AA} \sim 1.1$,
$\rho_c/\rho_0 \sim 0.416$. The glass transition locus was estimated
in \cite{sastryprl} using diffusivity data fitted to the VFT form
(Eq.(\ref{eq:vft})), and by thermodynamic means {\it via}
Eq. (\ref{eq:scdef}) (by the condition that $S_c$ vanishes at the
glass transition), where the total entropy of the liquid was obtained
by thermodynamic integration from the ideal gas reference state
(details in \cite{sastryprl}). The liquid-gas spinodal locus was also
obtained to study the relation between these two limits to the liquid
state. The two loci are seen to intersect at a finite $T$ leading to
the prediction of an ideal glass-gas mechanical instability locus at
low temperatures. Fig. 2 shows this ``phase diagram'' of the binary LJ
liquid, along with the locus of zero pressure (which shows that at low
enough density, the liquid is under tension at the simulated
temperatures). The figure also shows the $T$, $\rho$ points where
simulations have been performed (see \cite{sastryprl} for further
details).

The method of determining basin free energy and entropy (Eq.s
\ref{eq:fbasin},\ref{eq:sbdef}) from vibrational frequencies relies on
the assumption that the basins are harmonic. Some indirect tests of
this assumption have been presented recently
elsewhere\cite{sastrypcc}.  A direct test\cite{scala} would be to
thermally excite the inherent structures to excitation temperatures
$T_e$, and examine whether the energy of excitation above the inherent
structure energy obeys the harmonic expectation of $\Phi(T_e) -
\Phi_{min} = \ {3 \over 2} N k_B T$.  The result of this test is shown
in Fig. 3 for $\rho = 1.2 \rho_0$, where inherent structures are
selected from simulations at three different temperatures, and are
subjected to thermal excitations at $T_e$ that span a range around the
simulation temperature.  In all cases, the deviation from the harmonic
expectation is less than $4 \%$. Surprisingly, the deviations are
bigger at lower temperatures. The reason is most likely that very
closely related minima with negligible barriers exist at all energies
studied\cite{sastrynature}, and transitions between them occur even
for low $T_e$.  Nevertheless, the results in Fig. 3 show that in
practice, the harmonic approximation is quite valid for the binary LJ
liquid.

As described in the previous section, the classical calculation of the
basin entropies indicates that differences in basin entropy between
two basins is independent of temperature, and remain finite at zero
temperature. Since this is not true in real materials (where all
entropies vanish at zero temperature), one must examine to what extent
conclusions from such calculations apply to real systems. Thre
temperature range relevant to studying glass forming liquids is
clearly the range above the glass transition temperature. Further, the
relevant quantity for the analysis above was the difference of basin
entropies. Therefore, if the quantum mechanical basin entropy
differences correspond to a substantial fraction of the classical
basin entropy differences above the glass transition, we may conclude
that the classical approximation is reasonable. Fig. 4 shows classical
and quantum mechanical entropies obtained for basins sampled at two
different temperatures (one `high', one `low') at two densities. It is
clear (panels (a) and (d)) that the classical expression for the
entropy deviates strongly from the quantum analog at low $T$ (indeed,
it diverges to $-\infty$ as $T \rightarrow 0$) while approaching it at
high $T$. Similarly the quantum mechanical entropy differences
approach the constant classical value at high $T$ but decrease towards
zero at low $T$.  Nevertheless, at the glass transition temperature
$T_K$, the quantum mechanical difference is still about $67 \%$ of the
classical result. Since Argon units are used for the $A$ atoms, and
most glass formers are made of considerably bulkier molecules than
atomic Argon, we may conclude that the classical result must apply
well in most cases of interest.

\section{Summary and Conclusions}\label{secconc}
The discussion and results presented here indicate that the inherent
structure approach is very useful in addressing interesting questions
concerning the thermodynamics of glass forming liquids. It has been
shown that the thermodynamic estimate of the glass transition locus
agrees quite well with the one obtained from dynamic
data\cite{sastryprl}.  Fruther, quantities such as the fragility of a
liquid are reasonably well estimated by thermodynamic means, leading
to an expression of fragility in terms of parameters that quantify the
energy landscape of the liquid. Additional data are presented here
supporting the evaluation of basin entropies in a harmonic
approximation and the classical calculations typically employed in
such studies.

\begin{figure}[b]
\hbox to\hsize{\epsfxsize=1.0\hsize\hfil\epsfbox{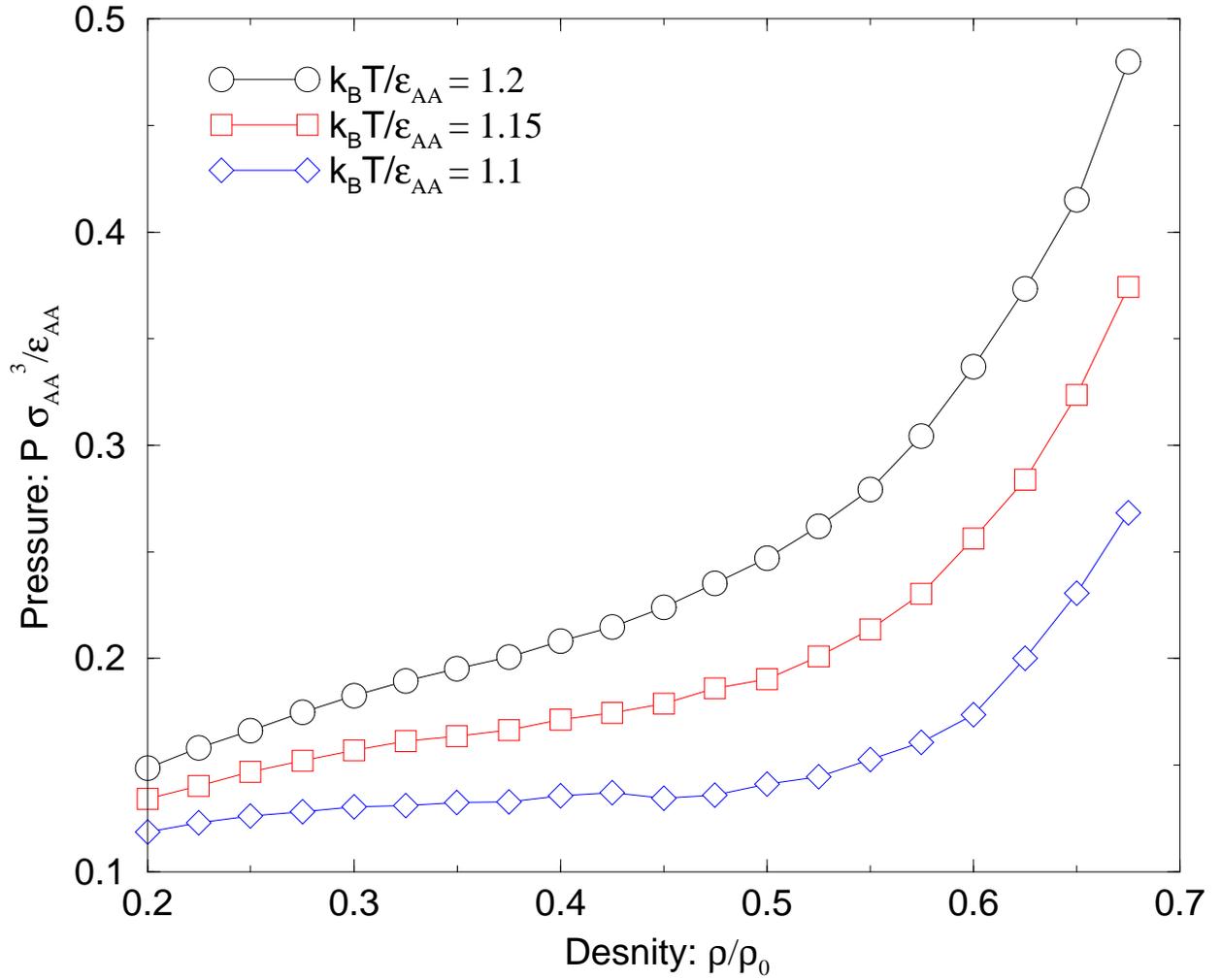}\hfil}
\caption{Isotherms at temperatures close to the liquid-gas critical
point, leading to an estiamate of the critical temperature and
density, $k_B T_c/\epsilon_{AA} \sim 1.1$, $\rho_c/\rho_0 \sim
0.416$.}
\label{fig1}
\end{figure}

\begin{figure}[b]
\hbox to\hsize{\epsfxsize=1.0\hsize\hfil\epsfbox{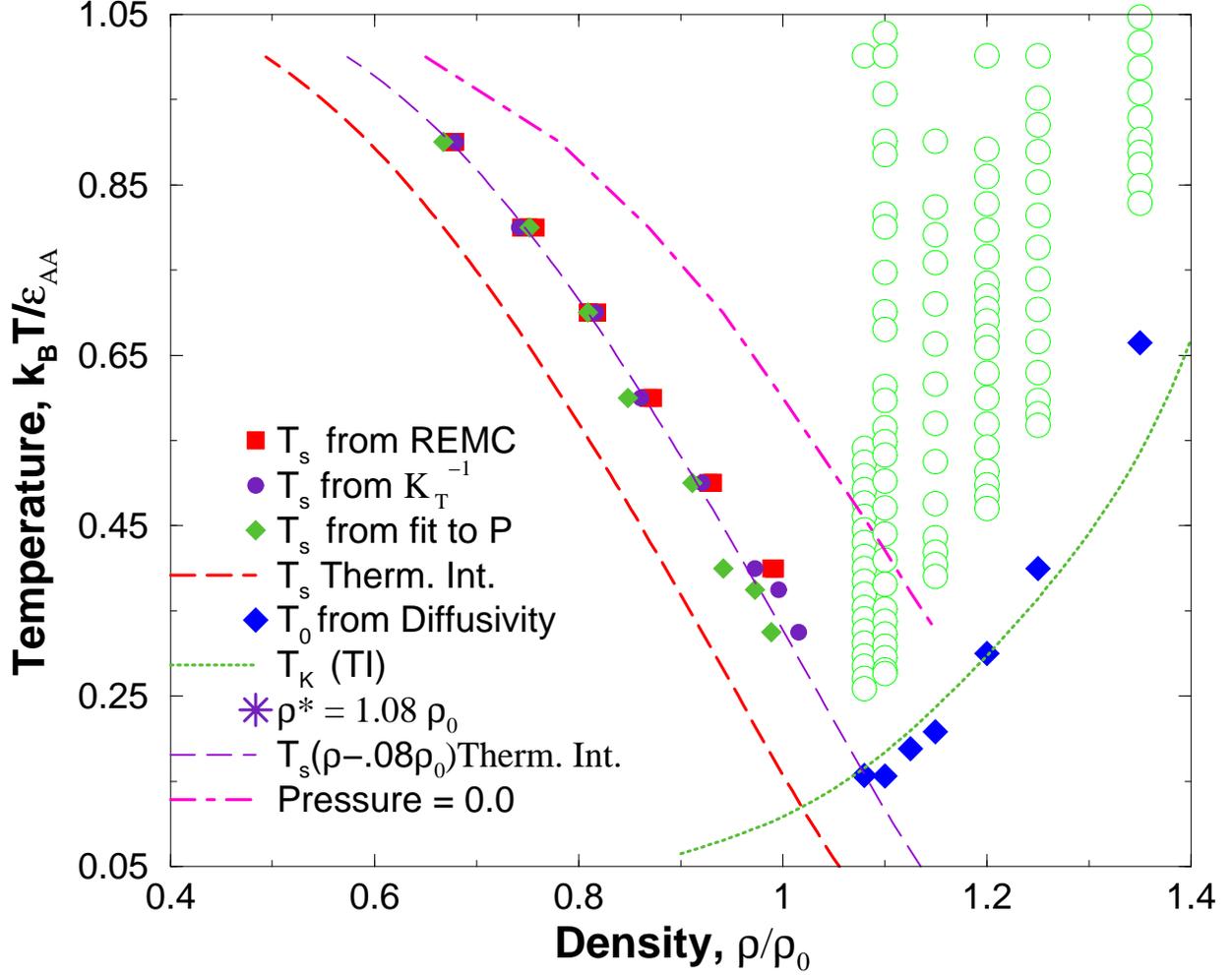}\hfil}
\caption{Liquid-gas spinodal obtained from (a) restricted ensemble
Monte Carlo simulations (REMC), (b) extrapolations of isothermal
compressibility $k_T$, (c) polynomial fits to isotherms obtained in
simulations, and (d) the empirical free energy obtained from
thermodynamic integration (`$T_s$ Therm. Int.'). The same curve is
also shown shifted in $\rho$ by $0.08 \rho_0$ (`$T_s (\rho-.08)$
Therm. Int.'). The glass transition locus obtained from (e) VFT fits
to diffusivity data, and (d) extrapolation of configurational entropy
to zero (`$T_{IG}$ Therm. Int.'). (*) marks the density $\rho^*$ where
inherent structure pressure is a minimum. The dot-dashed line
represents the locus of zero pressure. Open circles represent $T$,
$\rho$ values where simulations have been performed at high
densities.}
\label{fig2}
\end{figure}

\begin{figure}[b]
\hbox to\hsize{\epsfxsize=1.0\hsize\hfil\epsfbox{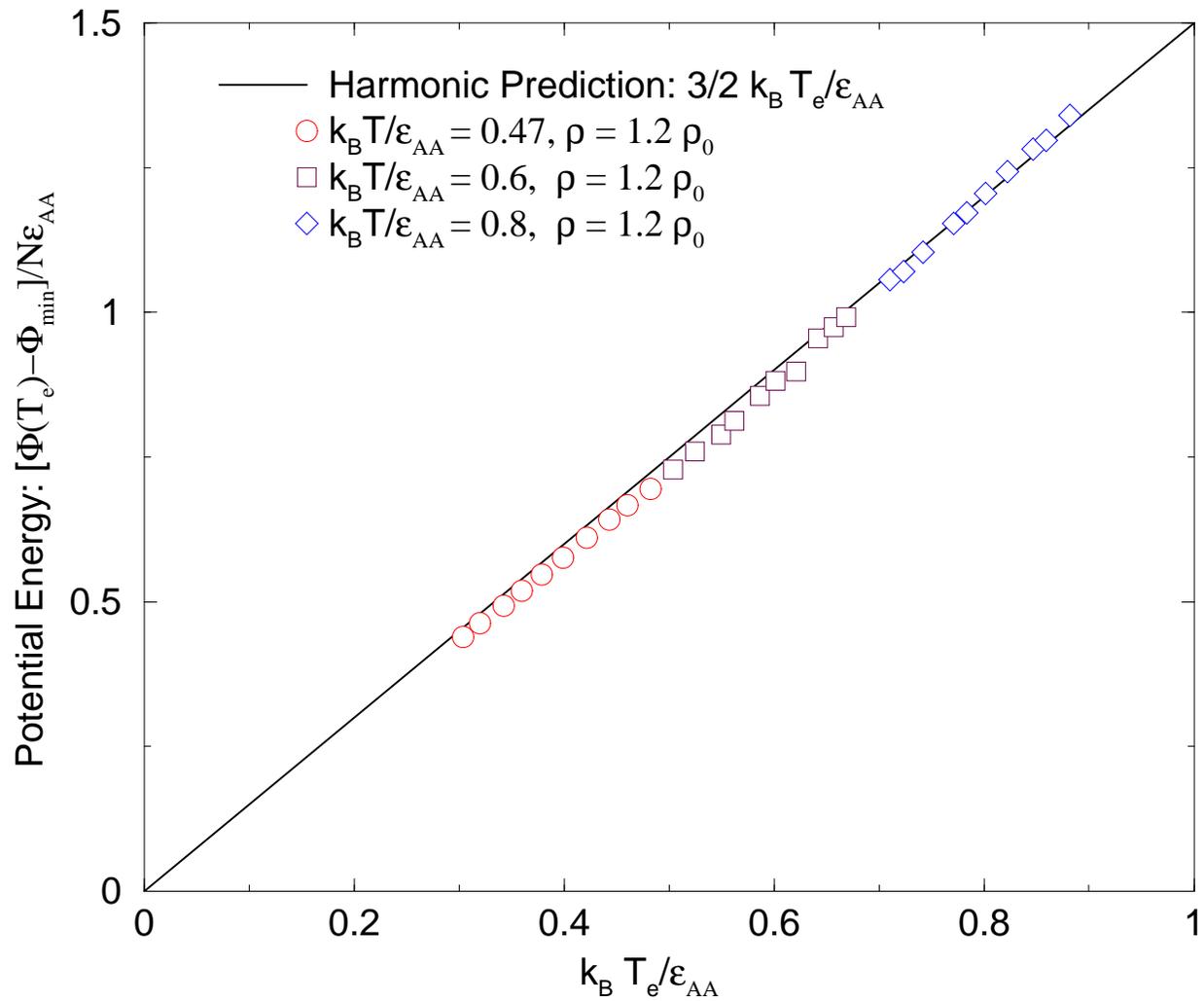}\hfil}
\caption{Test of the harmonic structure of minima. Energies shown are
obtained by thermally exciting inherent structures from runs at
different temperatures (shown in the legend), at `excitation
temperature's $T_e$ indicated on the x-axis and measuring the average
potential energy.}
\label{fig3}
\end{figure}

\begin{figure}[b]
\hbox to\hsize{\epsfxsize=1.0\hsize\hfil\epsfbox{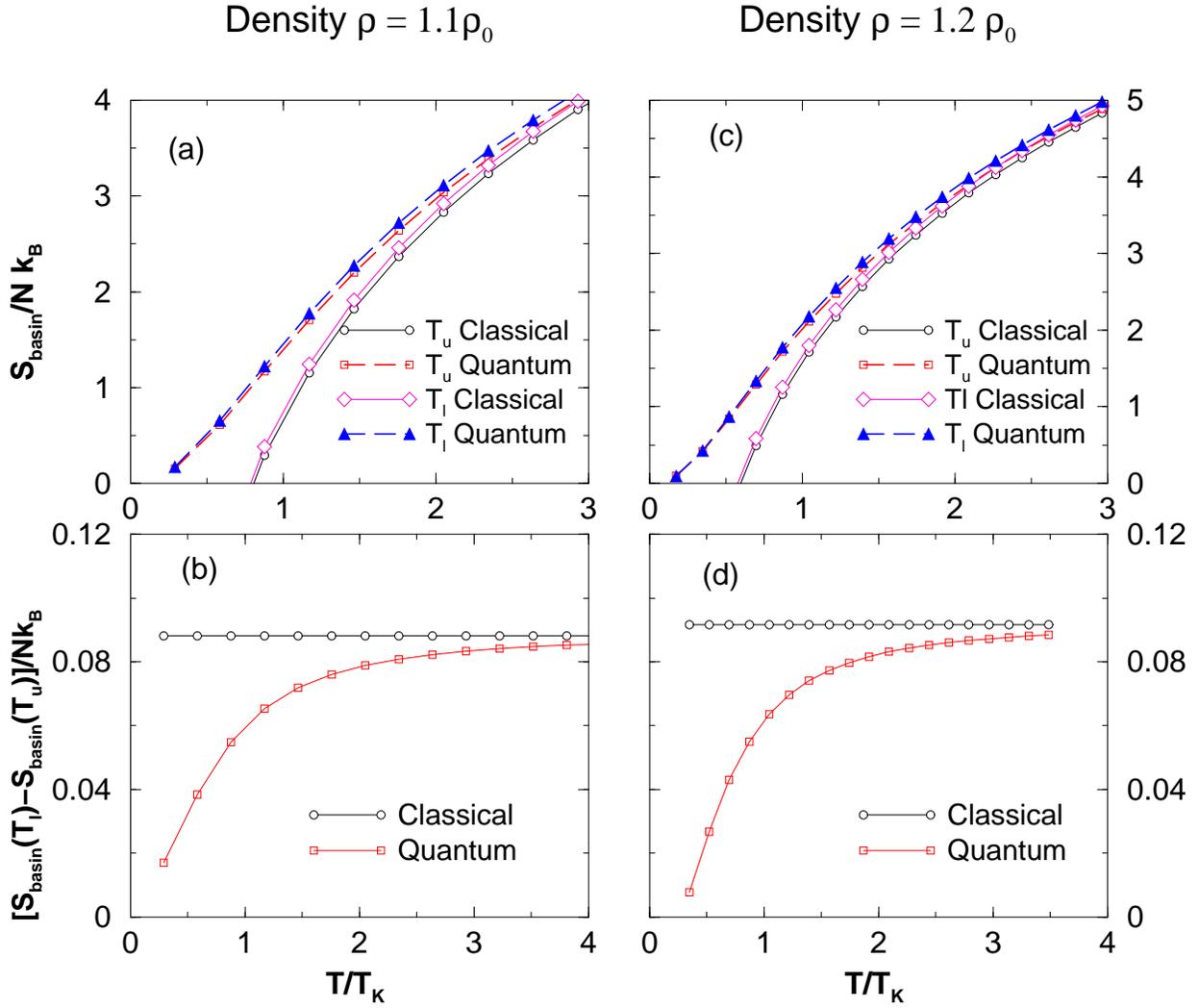}\hfil}
\caption{Basin entropies (panel (a) and (c)) and basin entropy
differences (panels (b) and (d)) at two densities of the liquid ($\rho
= 1.1 \rho_0$ for (a),(b), and $1.2 \rho_0$ for (c), (d)). At each
density, two basins sampled at a low (``$T_l$'') and high (``$T_u$'')
temperature are selected, for which the basin entropy {\it vs.} $T$ is
calculated both classically and quantum mechanically. For $1.1
\rho_0$, $k_B T_l/\epsilon_{AA} = 0.295$ $k_B T_u/\epsilon_{AA} =
0.6$, and for $1.2 \rho_0$, $k_B T_l/\epsilon_{AA} = 0.47$ $k_B
T_l/\epsilon_{AA} = 0.8$. Data shown indicates that in the quantum
mechanical case, the entropy differences to indeed vanish as $T
\rightarrow 0$, but at $T_K$, the quantum mechanical entropy
difference is roughly $67 \%$ of the classical result.}
\label{fig4}
\end{figure}

\end{document}